# Analyzing the Effects of COVID-19 Pandemic on the Energy Demand: the Case of Northern Italy [1]

Paolo Scarabaggio, Raffaele Carli, Massimo La Scala, Mariagrazia Dotoli

*Dept. of Electrical and Information Engineering*
*Polytechnic of Bari*
Bari, Italy
{paolo.scarabaggio, raffaele.carli, massimo.lascala, mariagrazia.dotoli}@poliba.it

**Abstract**

The COVID-19 crisis is profoundly influencing the global economic framework due to restrictive measures adopted by governments worldwide. Finding real-time data to correctly quantify this impact is very significant but not as straightforward. Nevertheless, an analysis of the power demand profiles provides insight into the overall economic trends. To accurately assess the change in energy consumption patterns, in this work we employ a multilayer feed-forward neural network that calculates an estimation of the aggregated power demand in the north of Italy, (i.e, in one of the European areas that were most affected by the pandemics) in the absence of the COVID-19 emergency. After assessing the forecasting model reliability, we compare the estimation with the ground truth data to quantify the variation in power consumption. Moreover, we correlate this variation with the change in mobility behaviors during the lockdown period by employing the Google mobility report data. From this unexpected and unprecedented situation, we obtain some intuition regarding the power system macro-structure and its relation with the overall people's mobility.

COVID-19; Lockdown; Power systems; Machine learning; Neural networks

## 1  Introduction

Currently, governments all around the world are fighting to contain the COVID-19 pandemic. Several countries, such as China, Italy, Germany, and France, adopted strict social distancing policies, which led in some cases to a total lockdown of their populations [1]. Traditionally, social and productive behaviors have a strong impact on electricity consumption patterns and therefore on the overall power system [2]. In particular, in the case of the COVID-19 pandemic in Italy, the consequence of such restrictive measures taken by the government are evident in terms of variations in the electricity consumption patterns [3]. Data on the actual effects of COVID-19 on economy are scarce, and hard to quantify due to the exceptional situation. Nevertheless, the economic framework is heavily dependent on the use of electricity [4,5], therefore, information on power demand variation may offer some insight for the real shock on the economic framework. Furthermore, this very peculiar situation can provide data useful for identifying load patterns and correlations useful for analysis in planning and operations [6].

The containment measures of the pandemic in Italy can be divided into two main phases, one following February 23, which mainly concerned Northern Italy, and a second one following March 09, including the more restrictive measures affecting the whole national territory. The first restrictive measures, which comprehend the closure of schools, universities, and bars and restaurants after 6 p.m., had limited effects on the contagion dynamics and as the crisis worsened the need for more severe restrictions motivated the second phase which turned into a total lockdown where all the nonessential production activities were shut down.

Preliminary analyses based on the data provided by Terna - the Italian transmission system operator (TSO) - shows that from March 9 onwards, there has been a strong decrease in energy consumption. However, the weather conditions in March contributed to a further slowdown in electricity consumption as a result of the increase in temperatures. Notwithstanding the cyclic pattern in power demand, a simple benchmark analysis to assess the effect of COVID19 is not enough, due to the different meteorological and socio-economic situations. Hence, machine learning (ML) techniques are more appropriate to correctly estimate the lockdown impact.

[1] This work received funding from the Italian University and Research Ministry under project RAFAEL (National Research Program, contract No. ARS01 00305) and from the European Union under the project OSMOSE (European Union's research project H2020 - call LCE-04-2017).

To the best of the authors' knowledge, only few works focus on ML-based analysis of the change in electricity patterns worldwide due to the shutdowns. Among the first contributions, Cicala [7] presents a preliminary analysis on the changes in the electricity consumption in Europe due to the spread of COVID-19. This analysis - aimed at providing a proxy for short-term changes in economic activity - was conducted until early April using a simple multivariate regression model. The results show that electricity consumption was reduced by 10% on average throughout the European Union in the week ending April 4 with respect to a baseline period in February. In particular, the percent change in Italy was about 23%, i.e., higher than the average since the COVID-19 restriction measures started earlier in the Italian country. Similarly, Narajewski and Ziel [8] more recently analyze the change in electricity demand profiles in various European countries, including Italy, due to the shutdowns. The authors use a high-dimensional regression technique based on [9], even thought external regressors like temperature are not taken into account. The analysis was conducted until April 13, mainly focusing on daily and weekly pattern. However, in the cited works [7,8] no details about the validation accuracy and the performance achieved by the used regression models are provided. In addition, in [7,8] the analysis is not extended to the entire shutdown period. We further remark that in the cited works [7,8] the analysis is limited to highlighting the variation in power consumption due to COVID-19; however, the variation results are not corroborated by further analysis aimed at detecting any inter-dependencies with other aspects.

Differently from the discussed related literature, in this work we analyze the effects on the aggregate energy demand in Northern Italy during the COVID-19 crisis until the end of May. We focus our analysis on a specific area and not on the whole country for two main reasons. On one hand, there are strong historical differences between the north and the south of Italy that led to a different socio-economic structure and therefore to different associated energy consumption patterns. On the other hand, the spread of COVID-19 and consequently of the restrictive measures had highly heterogeneous dynamics in the different regions.

In particular, we employ a multi-layer feed-forward neural network to characterize the aggregate energy demand in the north of Italy based on historical data. We use the network to correctly estimate the energy demand in the absence of the COVID-19 pandemic. Then, with the real-time data, we determine the effective energy demand variation caused by the lockdown. We employ the estimation to examine some of the critical interdependencies between socio-economic behaviors on the power infrastructure systems. In particular, we assume a correlation between mobility and energy demand variation by employing the Google mobility report data.

The paper remainder is organized as follows. Section II presents some preliminaries of the main ML techniques. Section III formulates the aggregated load demand forecast model and presents the estimation of the energy demand variation caused by the lockdown. Section IV analyzes the impacts of mobility on the aggregated load demand. Section V provides a summary with concluding remarks.

## 2 Preliminaries on Machine Learning

With the continuous growth in the availability of data, several techniques have been developed to extract unknown knowledge from novel sources of information. In particular, ML algorithms have been recognized as powerful tools for this scope, thanks to the mathematical models they rely on.

Let us introduce the basic notation in discriminative ML models [10]. In this class, we train the model with a large amount of known data and then we feed it with a new input trying to estimate the most probable output. More formally, we define a set of $n$ input variables $X = (X_1,...,X_n)$ usually called features, and a set of $r$ output variables $Y = (X_1,...,X_r)$ named target variables. However, in most of the practical ML applications, there are multiple features and a single target variable. We suppose that there is a vector-valued function $f(\cdot)$ which perfectly describes the relation between these two sets of variables.

The goal of ML algorithms is to guess the function $f(\cdot)$. The hypothesis of ML algorithms is a generic vector-valued function $h(\cdot)$ that returns an estimated value for $Y$ for any given input $X$. Thus, we have that:

$$Y = \hat{Y} + \epsilon = h(X) + \epsilon \tag{1}$$

where is the error associated with the guess and $\hat{Y}$ is the predicted output. We select the function $h(\cdot)$ based on a so-called training set, $\Xi$, of $m$ input vector examples. The aforementioned training set is composed of $m$ tuples: $(X^1, Y^1),...,(X^m, Y^m)$. For instance, the $i$th tuple $(X^i, Y^i)$ includes the single example record of the features and the relative target variables.

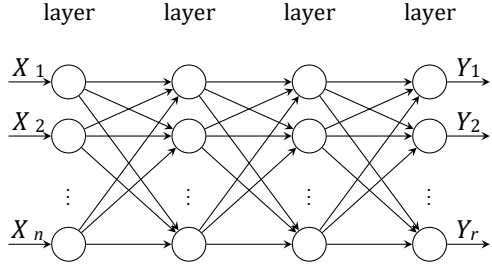

Figure 1: Example of a four-layer feed-forward neural network.

There are two main frameworks in which we attempt to guess the function $h(\cdot)$. In the first case, named supervised learning, we roughly know the relation within the data. In particular, we assume in advance the structure of the model and attempt barely to estimate its parameters. Conversely, in the second case, we simply feed the algorithm with the training set, without making any assumption on the structure of the function. By employing the minimum human supervision we are thus able to extract hidden patterns from data.

We finally remark that the first category includes two of the most important discriminative ML techniques, namely regression models (RM) and artificial neural networks (ANN), which will be both employed in the rest of this paper.

## 3 Aggregated Load Variation

Numerous researchers have considered the forecasting of electricity demand using a variety of modeling techniques. Models vary from simple linear reduction to human experience. However, ML techniques are recognized as one of the most powerful tools for this purpose [11].

In particular, ANNs are broadly used in energy forecasting and, in general, in pattern recognition applications due to their ability to learn complex nonlinear dynamics [12–14]. An ANN is composed of several computational nodes, the so-called neurons, which are connected together to solve complex problems by a specific training process. In detail, each neuron is trained for a specific application through a learning process by adjusting its synaptic weight.

In the ANN class, Multi-layer feed-forward neural networks (MFFNN) are the most commonly applied, due to their great versatility. These ANNs are characterized by a fully connected structure that consists of one input layer, one or more hidden layers, and one output layer. The overall network has a one-way structure: no loops are considered and all the neurons of the previous layer are fully connected to the neurons of the next layer. We show an example of these networks in Fig. 1.

In the energy sector, ANNs generally employ past energy demand records to predict future ones. However, the mere use of historical data is not sufficient, since energy demand is the result of complex interactions between meteorological and socio-economic factors. Therefore, a desirable model to predict energy demand requires, at least, the examination of the following variables or features:

- seasonal information such as month, day, hour and public holidays;

- meteorological conditions like temperature and precipitation; • social and economic data such as the gross

  domestic product;

- historical power demand records.

Load forecasting techniques such as ANNs are usually employed in the normal operation of power grids since they are able to predict with great accuracy the power that must be dispatched. All models rely on the fact that the power demand is highly cyclic and stable, and therefore extremely predictable in normal conditions. The impact of natural disasters or highly unpredictable events such as the COVID-19 pandemic is hard to estimate; however, forecasting techniques are still valuable. In particular, through a forecasting model, it is possible to estimate the energy needed in the normal operations, i.e., the aggregate load requested without the COVID-19 pandemic with the current meteorological and the expected socioeconomic factors. For the sake of clarity, we introduce the following three definitions used through the rest of this paper:
- *Estimation without COVID-19:* the energy demand calculated through the ANN forecasting model and therefore an estimation of the normal operation without the pandemic effects;
- *Real-time with COVID-19:* the real-time recorded energy demand by the Italian TSO (Terna), this value is obviously influenced by the pandemic;
- *Forecasting with COVID-19:* the day-ahead forecasting made by the Italian TSO (Terna), also this value is affected by the pandemic since the TSO takes into account the dynamical trend.

In the remaining part of the section, we select and train several ANNs employing historical data and we evaluate then through different metrics. We select the most reliable and we use it for the *Estimation without COVID-19*. Moreover, we compare the results with the real-time and the day-ahead energy demand.

## 3.1 Experimental Setup

In this study we analyze the aggregate energy demand in Northern Italy since it was the first area impacted by the virus crisis and consequently by the restrictive measures. Therefore, we gather electricity demand records in GWh from 2015 to 2020 for the North of Italy bidding zone from the Terna website [3], where the data are available based on a 15 minute sampling time.

As input data for the ANNs, several factors must be considered. To take into account the socio-economic factors we employ the gross domestic product (GDP) available on the world bank website on a four-month basis [15]. The environmental inputs of the model are the hourly precipitation, the hourly temperature, and the overall meteorological conditions. These data are available on different weather providers and for different meteorological stations, however, for the sake of simplicity, we use the data of the city of Milan [16]. Lastly, we include calendar data, for instance, by including variables that take into account public holidays, working days, and other important seasonal information. All the nonbinary variables are normalized with the z-score normalization technique [17].

Summing up, each tuple is composed of one target variable, i.e., the electricity demand in GWh for a specific 15 minutes sample, and ten features (temperature, precipitation, hour, day of the year, day of the week, month, public holiday, special day, seasonal data, and GDP).

Given the input data, we test several different MFFNNs by changing the number of hidden layers, the number of neurons per layer, and the training algorithms. In particular, we use the Levenberg-Marquardt back-propagation (LM) and Bayesian Regularization (BR) algorithms [18]. The first one minimizes the mean square error, while the latter minimizes a weighted sum of squared errors and squared weights.

We divided the sample data from 2015 to 2019 into three non-overlapping randomly selected subsets for training, validation, and testing, equal to 70%, 15%, and 15% of the overall data set, respectively. In particular, the training set was used to train the prediction models. The validation set was used for the tuning of the parameters of the selection of the prediction models, i.e., we validate the different network setups employing the validation set which examples are not been used to train the model. Moreover, after the selection of the best setup, we train again the model with the training and the validation set together. Lastly, we employ the testing data to finally evaluate the prediction model performance.

## 3.2 Network Design and Validation

To choose between the analyzed ANN configurations, we employ two different performance criteria, namely the mean absolute error (MAE) and mean squared error (MSE) defined as:

$$\text{MAE} = \frac{1}{M} \sum_{i=1}^{M} |Y^i - \hat{Y}^i| \qquad (2)$$

$$\text{MSE} = \frac{1}{M} \sum_{i=1}^{M} (Y^i - \hat{Y}^i)^2 \qquad (3)$$

where *M* is the size of the validation dataset. We remark that the MAE uniformly penalizes all errors, whereas the MSE heavily penalizes large errors. In Table 1 we show all the analyzed ANN configurations with the relative MAE and MSE calculated for the validation set. It is evident that the LM network with four hidden layers and 10 neurons per layer (i.e., Net05) exhibits the best performance. Therefore, this configuration will be used in the following analysis.

The aim of any ML algorithm is the ability to generalize and adapt to unseen data. In fact, the error of the model is usually lower on the training set than the validation set and this gap is referred to as the "generalization gap." The learning curves reported in Fig. 2 show that the selected network Net05 has a small validation error and a small generalization gap, thus indicating a well-fitted model.

Table 1: Comparison of performance achieved by the analyzed ANN configurations

| Name | Hidden layers | Neurons | Algorithm | MAE | MSE |
| --- | --- | --- | --- | --- | --- |
| Net01 | 2 | 5 | LM | 2.7472 | 14.3597 |
| Net02 | 2 | 10 | LM | 2.3261 | 11.9732 |
| Net03 | 3 | 5 | LM | 2.1247 | 13.1435 |
| Net04 | 3 | 10 | LM | 1.1164 | 3.0285 |
| Net05 | 4 | 10 | LM | 0.5311 | 1.0281 |
| Net06 | 2 | 5 | BR | 2.8540 | 15.7204 |
| Net07 | 2 | 10 | BR | 2.2142 | 10.6159 |
| Net08 | 3 | 5 | BR | 2.4196 | 13.0686 |
| Net09 | 3 | 10 | BR | 1.5994 | 5.2792 |
| Net10 | 4 | 10 | BR | 0.7896 | 1.5816 |

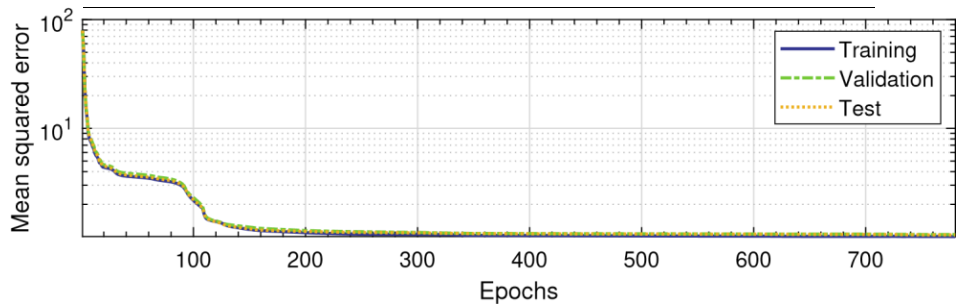

Figure 2: Training, validation, and test set MSE per epoch for the Net05.

Subsequently, we retrain the selected configuration by employing the training and validation set together, obtaining a final MAE on the test set of 0.3572. We calculate the MAE of the test set for the Terna forecasting system obtaining a value of 0.0943. We underline that the Terna forecasts are made the day ahead, therefore, they contain comprehensive and reliable information on the dynamical trends which takes into account the effects of lockdown and mobility restrictions. In our work, we disregard this information since we aim at estimate the energy consumption in the absence of the pandemic (i.e., *Estimation without COVID-19*), indeed, our goal is not to forecasts the real energy consumption but rather the demand that would have been supplied in a specific day in absence of lockdown and mobility restrictions. However, it is possible to improve the proposed model for forecasting purposes by employing a recurrent neural network to take into account the dynamical trends [19].

### 3.3 Numerical Results

As aforementioned, we select the Net05 as it shows the best performance. Therefore, in Fig. 3 we show the *Real-time with COVID-19*, the *Forecasting with COVID-19*, and the *Estimation without COVID-19*, i.e., the value calculated with the selected network.

The model shows its effectiveness in predicting the aggregate energy demand in January and February with a low error (Figs. 3a and 3b). At the same time, the model is able to detect the strong influence on the ground truth (i.e., the real-time data) produced by restrictive measures in March, April and May (Figs. 3c, 3d and 3e).

Moreover, it is worthwhile to show that the Terna forecasting system during the pandemic, i.e., the *Forecasting with COVID-19*, has a MAE of 0.4358, which indicates that the average error is roughly four times higher than during normal operations.

Several preliminary studies on the effect of COVID-19 on power systems were carried out by benchmarking the Real-time with COVID-19 value with the past years' data not taking into account other variables such as the weather conditions. These analyses [20] show a drop of more than 40% on the aggregate daily load. However, by employing the proposed model, we show in Fig. 4 that the reduction is lower than expected. The overall result shows that the maximum decrease of the aggregate load is 35%. Moreover, it is evident the restriction of February 23 did not affect the aggregate energy demand, while the restrictive measures of March 9 had the strongest impact in particular during the working days.

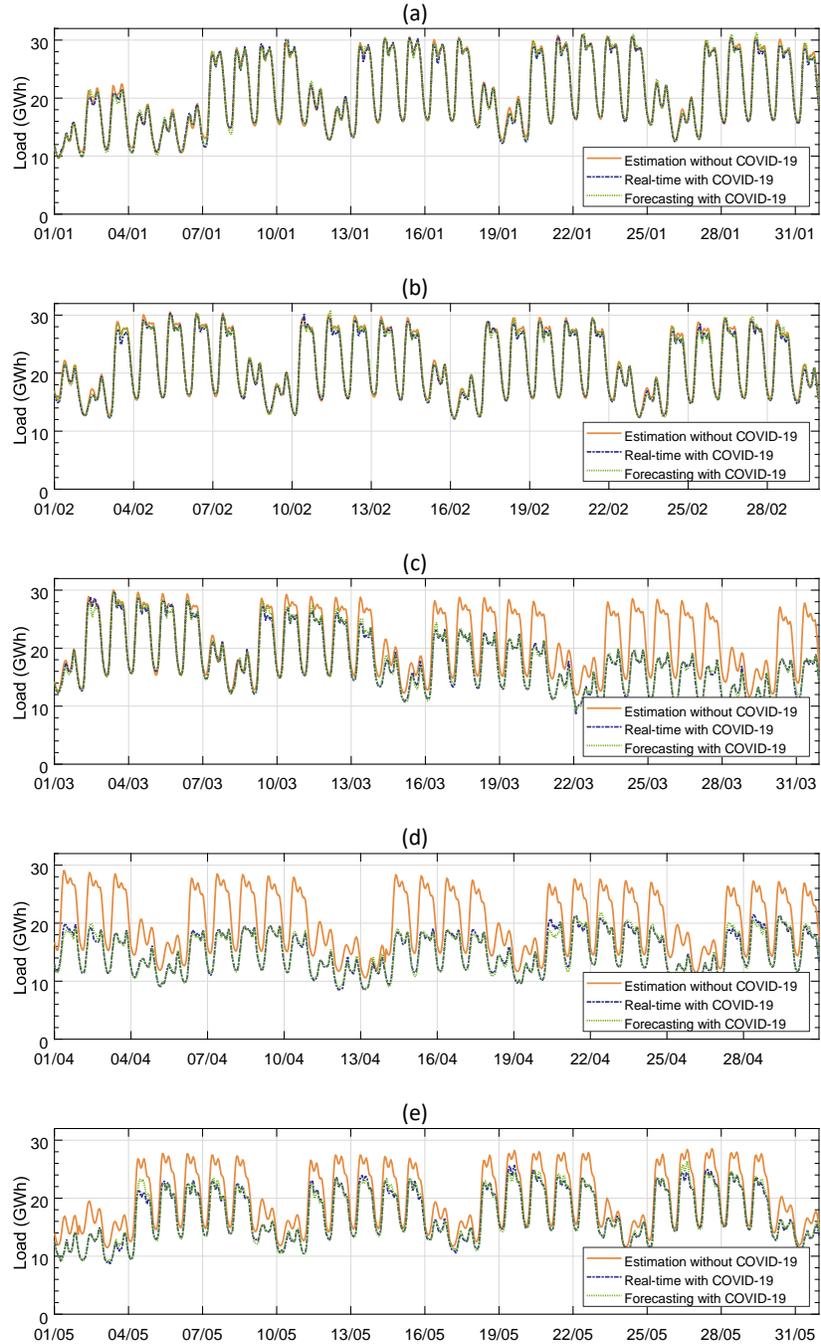

Figure 3: Aggregate load monthly profile: Estimation without COVID-19, Real-time with COVID-19 and Forecasting with COVID-19 for January (a), February (b), March (c), April (d) and May (e).

## 4   Impact of Mobility on the Aggregated Load

Electricity and electrical energy services are essential to the proper functioning of modern life, in fact, the power system infrastructure fulfills this critical role by ensuring the continuous availability of electrical energy services. As society and communities evolve, the energy demand and social behavior become more and more interconnected. Therefore, a successful investigation of the effect of COVID-19 on power infrastructure must consider these links and interdependencies. In this study we aim at examining some of these crucial inter-dependencies on the power infrastructure

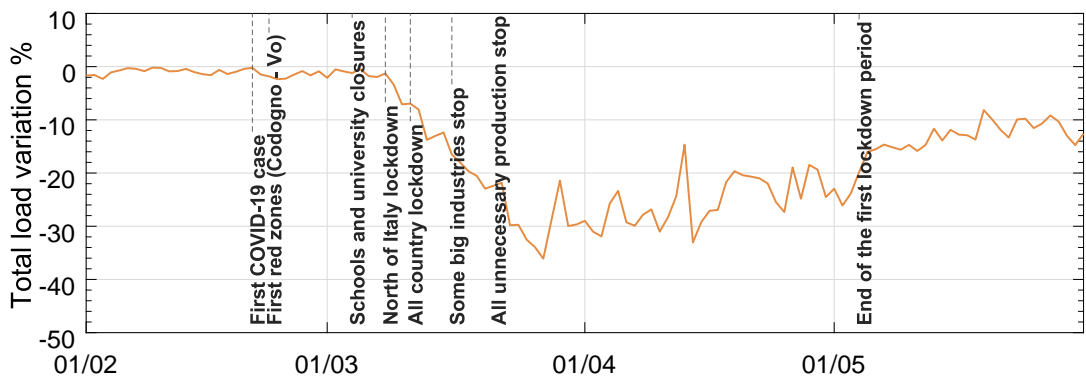

Figure 4: Percentage variation of the daily energy load due to COVID-19 pandemic.

by employing the Google mobility reports [21].

## 4.1 Google Mobility Data

The Google mobility reports show how visits and length of stay at different places change compared to a baseline. Changes for each day are compared to a baseline value for that day of the week. This dataset is based on data from users who have turned on the location history for their account; hence, the data represents a sample of all Google users. In particular, the mobility trends are divided into different categories:

- *Transit stations:* subway, bus, and train stations;
- *Retail & recreation:* restaurants, cafes, shopping centers, museums, libraries, and cinema;
- *Workplaces*;
- *Residential*;
- *Grocery & pharmacy:* grocery markets, food warehouses, specialty food shops, and pharmacies;
- *Parks:* local parks, national parks, public beaches, marinas, and public gardens.

In our analysis, we employ only the first four mobility categories since they are recognized as the most influential on the energy demand consumption. Moreover, we select only the data related to northern Italy. In Fig. 5 we show the mobility data from February 15 for the four considered categories. Similarly to the energy demand data, from Fig. 5 it is apparent that only after the total lockdown of March 9 we have a strong variation in social behaviors.

## 4.2 Regression Analysis and Results

Employing the neural network of Section III we first calculate the percentage variation of the daily energy demand with respect to a baseline: indeed, this value is somehow related to the daily variation of mobility in the Google reports. Subsequently, due to the limited number of samples, we use a simple linear regression technique to model the relation between these variables.

In particular, multivariate linear regression is a widely used approach to examine relationships between quantitative data. The model assumes that a dependent variable (i.e., the target variable) can be estimated by a linear combination of some independent variables (i.e., the features). Moreover, it is possible to include additional features in the regression model. For instance, it is possible to multiply two different features or elevate to the square or cube a single variable. This process is useful when the relation between the dependent variables and the independent variable cannot be properly described by a linear combination between features. For example, it is possible to formulate a second-order multivariate linear regression model with *n* input as:

$$Y = \beta_0 + \beta_1 X_1 + \beta_{11}(X_1)^2 + \ldots + \beta_n X_n + \beta_{nn}(X_n)^2 + \varepsilon \tag{4}$$

where $Y$ is the dependent variable, $X_i$ are the independent variables, $\beta_0$ is the constant parameter of the model, and $\beta_i$ and $\beta_{ij}$ are the remaining coefficients. Note that in (4) $(X_n)^2$ is an example of extra feature created by squaring a single independent feature. The order of the polynomial model, i.e., the order made by adding additional features, should be kept as low as possible because an arbitrary fitting of higher-order polynomials provides a "good" fit to the data, however, such models neither enhance the understanding of the unknown function nor are a good predictor. One possible approach is to successively fit the models in increasing order and test the significance of regression coefficients at each step of fitting.

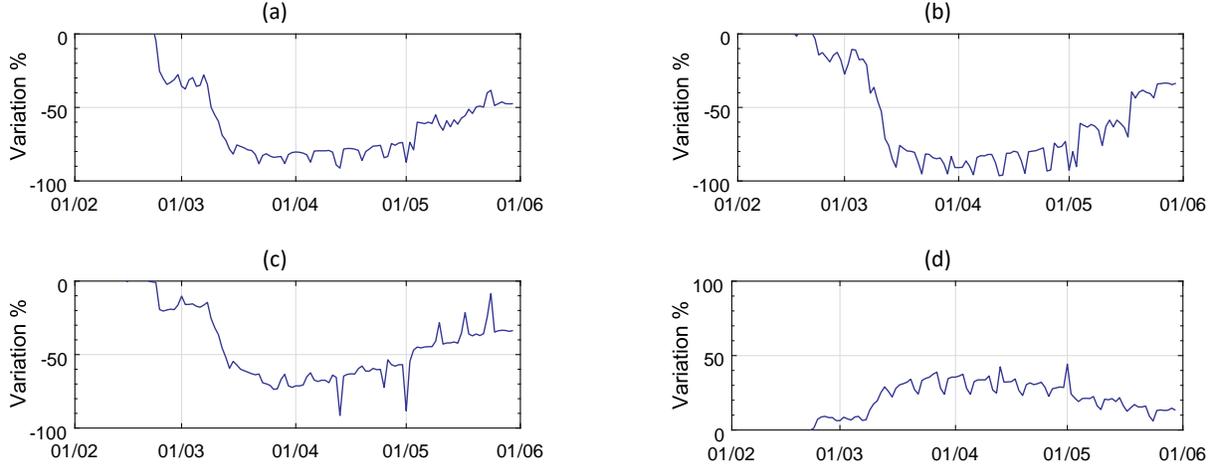

Figure 5: Google mobility reports: mobility variation for transit stations (a), retail & recreation (b), workplaces (c), and residential (d).

Since the purpose of this work is assessing the influence of mobility on power systems, we use the Google mobility reports values as the input variables and the energy demand variation as the output variable. The research further narrows the examination of this relation by checking whether it is best described by a linear, quadratic, or cubic model.

Accordingly, let us employ as a goodness-of-fit measure for regression models the MAE and the $R^2$ indicator. The former indicates the percentage of the variance in the dependent variable that the independent variables explain collectively. The $R^2$ measures how well the regression predictions approximate the real data points, and is defined as:

$$R^2 = \sum_{i=1}^{M}(\hat{Y}^i - \overline{Y})^2 / \sum_{i=1}^{M}(Y^i - \overline{Y})^2 \tag{5}$$

where $\overline{Y}$ is the mean of the observed value. Due to the small number of samples, a standard partitioning between the training and validation set would not be fruitful, therefore, let us employ the so-called leave-one-out cross-validation (LOOCV) [22]. In Table 2 we show the MAE and the $R^2$ respectively for the whole data and the LOOCV procedure when it is assumed a linear, quadratic, or a cubic model respectively. Moreover, in Fig. 6 we show the goodness-of-fit plots to offer a visual comparison between the data and fitted distributions, i.e., the observed energy variation with respect to the predicted energy variation of the regression model. Both results show that the quadratic model provides a good fitting.

To further prove this, in Fig. 7 we show the total load variation predicted by the above defined quadratic model and the related observed values as a function of the mobility variation for the four mobility categories. From Fig. 7 the quadratic relation between the total load variation and the mobility variation is evident in all the considered categories.

Finally, from the above computed regression coefficients we estimate the sensitivity factors of the total load variation with respect to the mobility variation for the four categories in different phases of the pandemic outbreak. For instance, we estimate the sensitivity factor for the transit station category by taking the mobility value for a specific day and modifying only the transit mobility variation by a given value ΔMobilityVar$_{Transit}$. Afterward, we observe how this change influences the total load variation ΔLoadVar. The sensitivity factors for a specific days in the prelockdown, full lockdown and post-lockdown phase are [0.034, 0.025, 0.159, −0.413], [0.046, 0.008, 0.003, −1.190] and [0.015, 0.02, 0.055, −0.536],

respectively. In all days, the residential sensitivity factor is negative, indicating that if the residential mobility increases, the total load decreases. Indeed, the larger number of people staying at home, the higher increment of the residential load. Nevertheless, the lower energy consumption caused by interdependencies with the other mobility categories generates an overall decrease in the aggregate energy consumption. From these factors, it is also evident that in the Italian power system the impact of the residential load is higher during the lockdown and Table 2: Comparison of the performance achieved by the analyzed regression models the post-lockdown phases. Conversely, the workplaces' sensitivity factor during the full lockdown phase is low: this indicates that the power system has an industrial baseload that remains almost constant during the lockdown period.

|           | MAE    | $R_2$   | LOOCV MAE | LOOCV $R^2$ |
|-----------|--------|--------|-----------|-------------|
| Linear    | 0.6247 | 0.9087 | 0.6690    | 0.8940      |
| Quadratic | 0.0835 | 0.9943 | 0.1001    | 0.9925      |
| Cubic     | 0.0836 | 0.9953 | 0.1242    | 0.9912      |

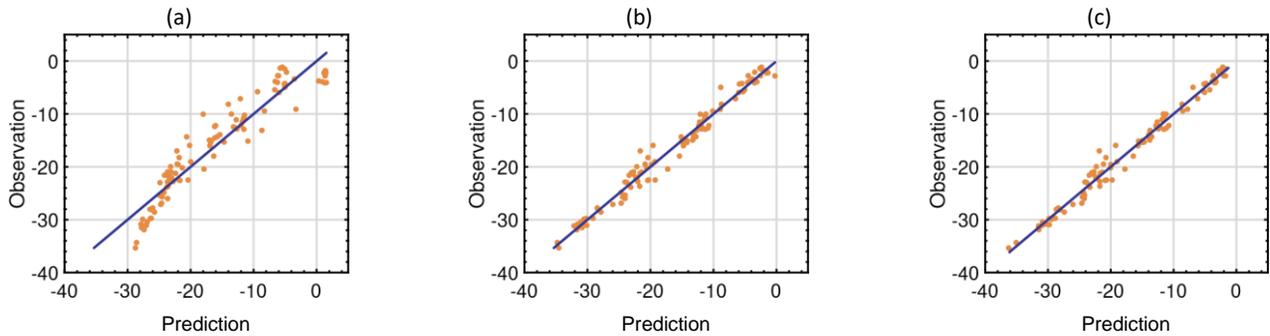

Figure 6: Goodness-of-fit plots for the Linear (a), Quadratic (b), and Cubic (c) model.

## 5 Conclusions

This paper presents an approach to quantify the impact of the COVID-19 restrictive measures on power systems and, shows the characteristic inter-dependencies between power systems and social behaviors (such as mobility patterns). The paper shows that most of these attitudes may have strong associations in energy demand. Moreover, the relations identified in this paper can also provide a basis for future resilience assessments and resilience designs of power systems.

Future research will be focused on additional correlation factors. In particular, we will estimate how a deeper penetration of smart working and smart mobility affects the energy consumption patterns.

## References

[1] N. Ferguson, D. Laydon, G. Nedjati-Gilani, N. Imai, K. Ainslie, M. Baguelin, S. Bhatia, A. Boonyasiri, Z. Cucunubá, and G. Cuomo-Dannenburg, "Impact of non-pharmaceutical interventions to reduce covid-19 mortality and healthcare demand." *Preprint at Spiral*, 2020.

[2] J. Chontanawat, L. C. Hunt, and R. Pierse, "Does energy consumption cause economic growth?: Evidence from a systematic study of over 100 countries," *Journal of policy modeling*, vol. 30, no. 2, pp. 209–220, 2008.

[3] Terna.it. 2020. [Online]. Available: www.terna.it

[4] R. Baldwin and B. Weder, *Economics in the Time of COVID-19*. CEPR Press, 2020.

[5] R. Carli, M. Dotoli, and R. Pellegrino, "Multi-criteria decision-making for sustainable metropolitan cities assessment," *Journal of environmental management*, vol. 226, pp. 46–61, 2018.

[6] S. Abbasi, H. Abdi, S. Bruno, and M. La Scala, "Transmission network expansion planning considering load correlation using unscented transformation," *Int. Journal of Electrical Power & Energy Systems*, vol. 103, pp. 12–20, 2018.

[7] S. Cicala, "Early economic impacts of covid-19 in europe: A view from the grid," Tech. rep. University of Chicago, Tech. Rep., 2020.

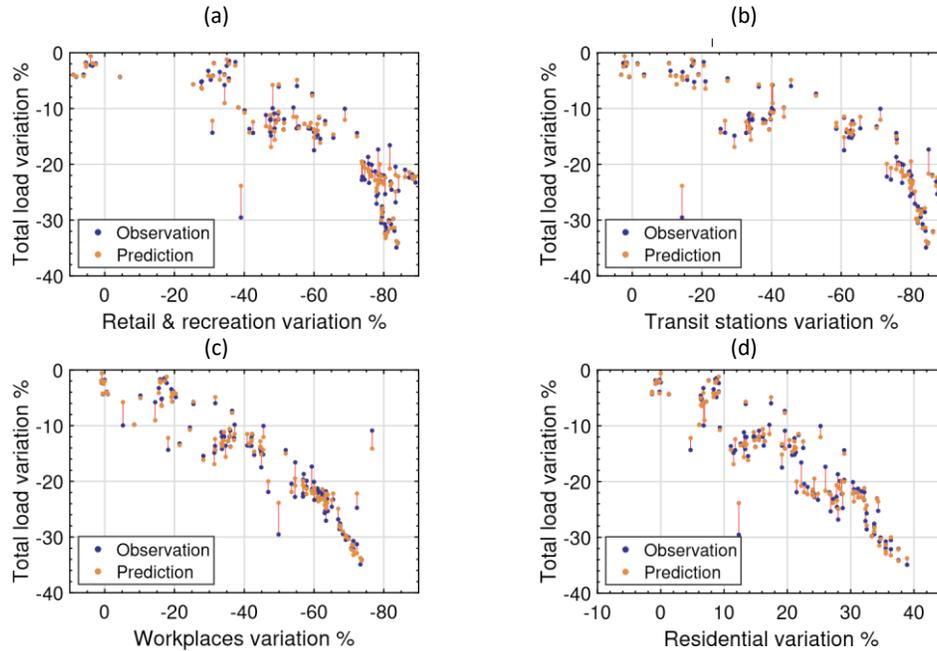

Figure 7: Predicted and observed aggregate load variation with respect to the mobility variation for transit stations (a), retail & recreation (b), workplaces (c), and residential (d).

[8] M. Narajewski and F. Ziel, "Changes in electricity demand pattern in europe due to covid-19 shutdowns," *preprint arXiv:2004.14864*, 2020.
[9] F. Ziel and B. Liu, "Lasso estimation for gefcom2014 probabilistic electric load forecasting," *Int. Journal of Forecasting*, vol. 32, no. 3, pp. 1029–1037, 2016.
[10] E. Alpaydin, *Introduction to machine learning*. MIT press, 2020.
[11] R. Blonbou, "Very short-term wind power forecasting with neural networks and adaptive bayesian learning," *Renewable Energy*, vol. 36, no. 3, pp. 1118–1124, 2011.
[12] G. Oğcu, O. F. Demirel, and S. Zaim, "Forecasting electricity consumption with neural networks and support vector regression," *Procedia-Social and Behavioral Sciences*, vol. 58, pp. 1576–1585, 2012.
[13] H. Hamedmoghadam, N. Joorabloo, and M. Jalili, "Australia's long-term electricity demand forecasting using deep neural networks," *preprint arXiv:1801.02148*, 2018.
[14] P.-H. Kuo and C.-J. Huang, "A high precision artificial neural networks model for short-term energy load forecasting," *Energies*, vol. 11, no. 1, p. 213, 2018.
[15] World bank open data. 2020. [Online]. Available: data.worldbank.org
[16] Openweathermap. [Online]. Available: www.openweathermap.org
[17] A. Jain, K. Nandakumar, and A. Ross, "Score normalization in multimodal biometric systems," *Pattern recognition*, vol. 38, no. 12, pp. 2270–2285, 2005.
[18] A. Payal, C. Rai, and B. Reddy, "Comparative analysis of bayesian regularization and levenberg-marquardt training algorithm for localization in wireless sensor network," in *2013 15th Int. Conference on Advanced Communications Technology*. IEEE, 2013, pp. 191–194.
[19] W. Kong, Z. Y. Dong, Y. Jia, D. J. Hill, Y. Xu, and Y. Zhang, "Short-term residential load forecasting based on lstm recurrent neural network," *IEEE Trans. on Smart Grid*, vol. 10, no. 1, pp. 841–851, 2017.
[20] Covid-19 effects. [Online]. Available: www.bruegel.org/publications/datasets/bruegel-electricity-tracker-of-covid-19-lockdown-effects
[21] Google llc. 2020. covid-19 community mobility reports. [Online]. Available: www.google.com/covid19/mobility/
[22] A. Vehtari, A. Gelman, and J. Gabry, "Practical bayesian model evaluation using leave-one-out cross-validation and waic," *Statistics and computing*, vol. 27, no. 5, pp. 1413–1432, 2017.